\documentclass{article}
\usepackage[preprint]{spconfa4}
\usepackage{amsmath,graphicx}
\usepackage{booktabs}
\usepackage[sort&compress,numbers]{natbib}

\copyrightnotice{978-1-6654-6867-1/22/\$31.00~\copyright2022 IEEE}
                   
\title{Speaker-conditioning Single-channel Target Speaker Extraction using Conformer-based Architectures}
\name{{Ragini Sinha$^1$, Marvin Tammen$^2$, Christian Rollwage$^1$, Simon Doclo${^{1,2}}$}}
\address{$^1$Fraunhofer Institute for Digital Media Technology IDMT,\\ Oldenburg Branch for Hearing, Speech and Audio Technology HSA, Germany\\
$^2$Department of Medical Physics and Acoustics and Cluster of Excellence
Hearing4all,\\ University of Oldenburg, Germany\\
Email:{ragini.sinha@idmt.fraunhofer.de}}
\begin{document}
%
\maketitle
\begin{abstract}
Target speaker extraction aims at extracting the target speaker from a mixture of multiple speakers exploiting auxiliary information about the target speaker. In this paper, we consider a complete time-domain target speaker extraction system consisting of a speaker embedder network and a speaker separator network which are jointly trained in an end-to-end learning process. We propose two different architectures for the speaker separator network which are based on the convolutional augmented transformer (conformer). The first architecture uses stacks of conformer and external feed-forward blocks (Conformer-FFN), while the second architecture uses stacks of temporal convolutional network (TCN) and conformer blocks (TCN-Conformer). Experimental results for 2-speaker mixtures, 3-speaker mixtures, and noisy mixtures of 2-speakers show that among the proposed separator networks, the TCN-Conformer significantly improves the target speaker extraction performance compared to the Conformer-FFN and a TCN-based baseline system.
\end{abstract}
\section{Introduction} \label{sec:intro}
Recent advances in deep-neural networks (DNNs) have greatly improved the performance of speaker extraction systems, yet it remains a challenge when multiple speakers and background noise are present in the mixture. One possibility is to first extract all individual speakers from the mixture using blind source separation \cite{makino2018audio,vincent2018audio,hershey2016deep,kolbaek2017multitalker,luo2019conv,luo2020dual,subakan2021attention} and then select the target speaker from the extracted speakers. Alternatively, a speaker-conditioning target speaker extraction approach can be utilized, which requires auxiliary information about the target speaker to guide the DNN towards directly extracting the target speaker from the mixture \cite{wang2018voicefilter,li2020atss,zhang2020x,ge2020spex+,vzmolikova2019speakerbeam,delcroix2020improving,ephrat2018looking,afouras2018conversation,gu2019neural,li2020listen,ceolini2020brain,delcroix2021speaker}. Commonly used auxiliary information includes reference speech of the target speaker \cite{wang2018voicefilter,li2020atss,zhang2020x,ge2020spex+,vzmolikova2019speakerbeam,delcroix2020improving}, visual information \cite{ephrat2018looking,afouras2018conversation}, directional information \cite{gu2019neural}, brain signal \cite{ceolini2020brain}, or speech activity information of the target speaker \cite{delcroix2021speaker}. In this paper, we focus on single-channel target speaker extraction using reference speech as auxiliary information.

\begin{figure}[tb]
	\begin{minipage}[b]{1.0\linewidth}
		\centering
		\centerline{\includegraphics[width=6.3cm]{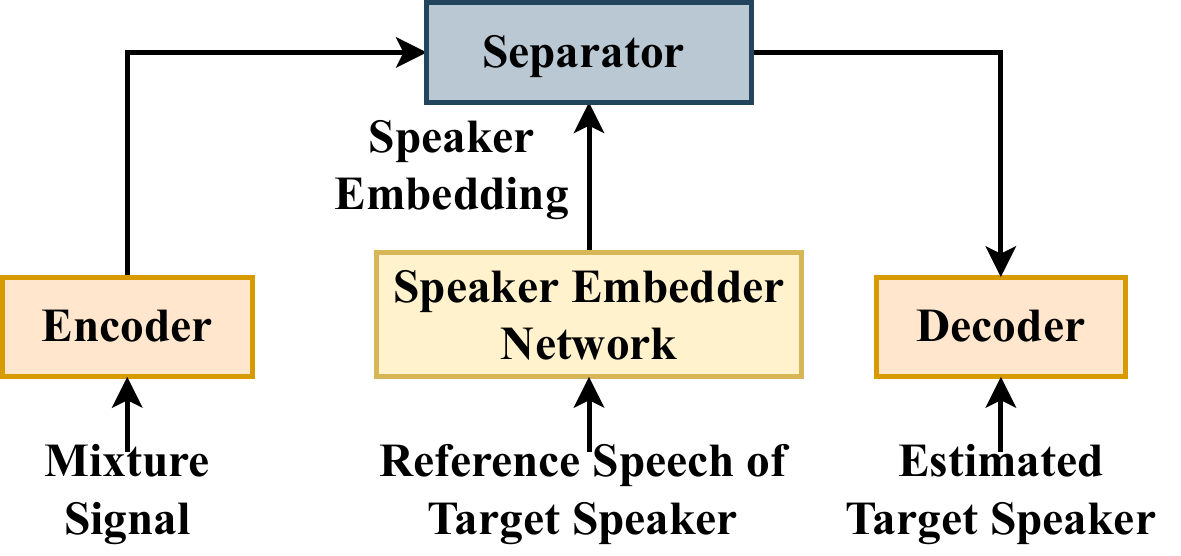}}
	\end{minipage}
	\caption{Block diagram of a time-domain speaker-conditioning target speaker extraction system.}
	\label{fig:ttse}
\end{figure}
A typical speaker-conditioning target speaker extraction system consists of a speaker embedder network and a speaker separator network. The speaker embedder network generates an embedding from the reference speech of the target speaker, while the speaker separator network aims at estimating the target speaker from the mixture guided by the speaker embedding. The speaker embedder and separator networks can be trained either separately \cite{wang2018voicefilter, li2020atss, zhang2020x} or jointly \cite{ge2020spex+} to perform the speaker extraction. In \cite{wang2018voicefilter, li2020atss, zhang2020x,ge2020spex+} different combinations of network architectures have been investigated for the embedder and separator networks. An LSTM-based embedder network was used in combination with a CNN-LSTM based separator network \cite{wang2018voicefilter} or a TCN-based separator network \cite{zhang2020x}. In \cite{li2020atss, ge2020spex+} ResNet-based architectures were used for the embedder network in combination with an attention-based separator network \cite{li2020atss} or a TCN-based separator network \cite{ge2020spex+}. The separator networks in \cite{wang2018voicefilter, li2020atss} perform the speaker extraction in the frequency-domain and inherently suffer from phase estimation issues, whereas the separator networks in \cite{ge2020spex+, zhang2020x} avoid phase estimation issues by performing the speaker extraction in the time-domain. In this paper, we consider complete time-domain target speaker extraction systems, where the embedder and separator networks are jointly trained.

Aiming at improving the performance of the TCN-based baseline system \cite{ge2020spex+} for target speaker extraction, in this paper we propose two different architectures for the speaker separator network based on the convolutional augmented transformer (conformer) \cite{gulati2020conformer}, which has shown its efficiency in capturing global feature information using attention and local feature information using convolution. The first proposed architecture is based on stacking conformer blocks and external feed-forward blocks (Conformer-FFN), whereas the second proposed architecture is based on stacking TCN blocks and conformer blocks (TCN-Conformer). Experimental results for 2-speaker mixtures, 3-speaker mixtures, and noisy mixtures of 2-speakers simulated using the WSJ0 dataset \cite{hershey2016deep} and WHAM dataset \cite{wichern2019wham} show that the proposed TCN-Conformer system significantly improves the speaker extraction performance for all mixture types compared to the TCN-based baseline system \cite{ge2020spex+} and proposed Conformer-FFN system in terms of scale-invariant signal-to-distortion ratio (SI-SDR).

\section{Time-domain Target Speaker Extraction System} \label{sec:tse}
A complete time-domain speaker-conditioning target speaker extraction system consists of a speaker embedder network and a speaker separator network both performing in the time-domain. The speaker separator network consists of three processing stages (see Fig.~\ref{fig:ttse}): encoder, separator, and decoder. The encoder transforms segments of the mixture signal to their intermediate feature representation, while the decoder reconstructs the target speaker signal from the masked encoded features. The multiplicative masks are estimated from the encoded features guided by the speaker embedding which is obtained from the reference speech of the target speaker using the speaker embedder network.  

In \cite{ge2020spex+}, the first stage of the speaker separator network uses a multi-scale encoder with three different filter lengths (low, medium, and high) to obtain different intermediate feature representations. The different filter lengths help in obtaining different temporal resolutions. The masks for the target speaker are estimated from the concatenation of the speaker embedding and the multi-scale encoded features using a TCN-based separator network for each filter length. The target speaker signal is reconstructed from the masked encoded features using a multi-scale decoder for each filter length. A multi-task learning procedure is utilized to train the speaker embedder and speaker separator networks jointly, where the separator network is optimized for speaker extraction using a multi-scale scale-invariant signal-to-noise ratio (SI-SNR), while the embedder network is optimized for speaker identification. The multi-scale SI-SNR is the weighted sum of the SI-SNRs estimated for all target signals obtained using multi-scale decoding. Similarly to \cite{ge2020spex+}, in this paper we consider the same ResNet-based embedder network architecture, but instead of using a TCN-based separator network we propose two different architectures based on the conformer. 

\section{Conformer-based Architectures} \label{sec:sep}
In this section, we discuss the proposed conformer-based architectures for the speaker separator network. The conformer \cite{gulati2020conformer} is capable of incorporating both local context as well as global context features. In Section \ref{sec:conf}, we discuss the Conformer-FFN architecture combining conformer blocks with feed-forward blocks to exploit local and global context features with the help of attention and convolutional operations. In Section \ref{sec:tcn-conf}, we discuss the TCN-Conformer architecture combining TCN blocks with Conformer blocks to exploit local and global context features.

\subsection{Conformer-FFN architecture}\label{sec:conf}
\begin{figure}[tb]
	\begin{minipage}[b]{1.0\linewidth}
		\centering
		\centerline{\includegraphics[width=9.2cm]{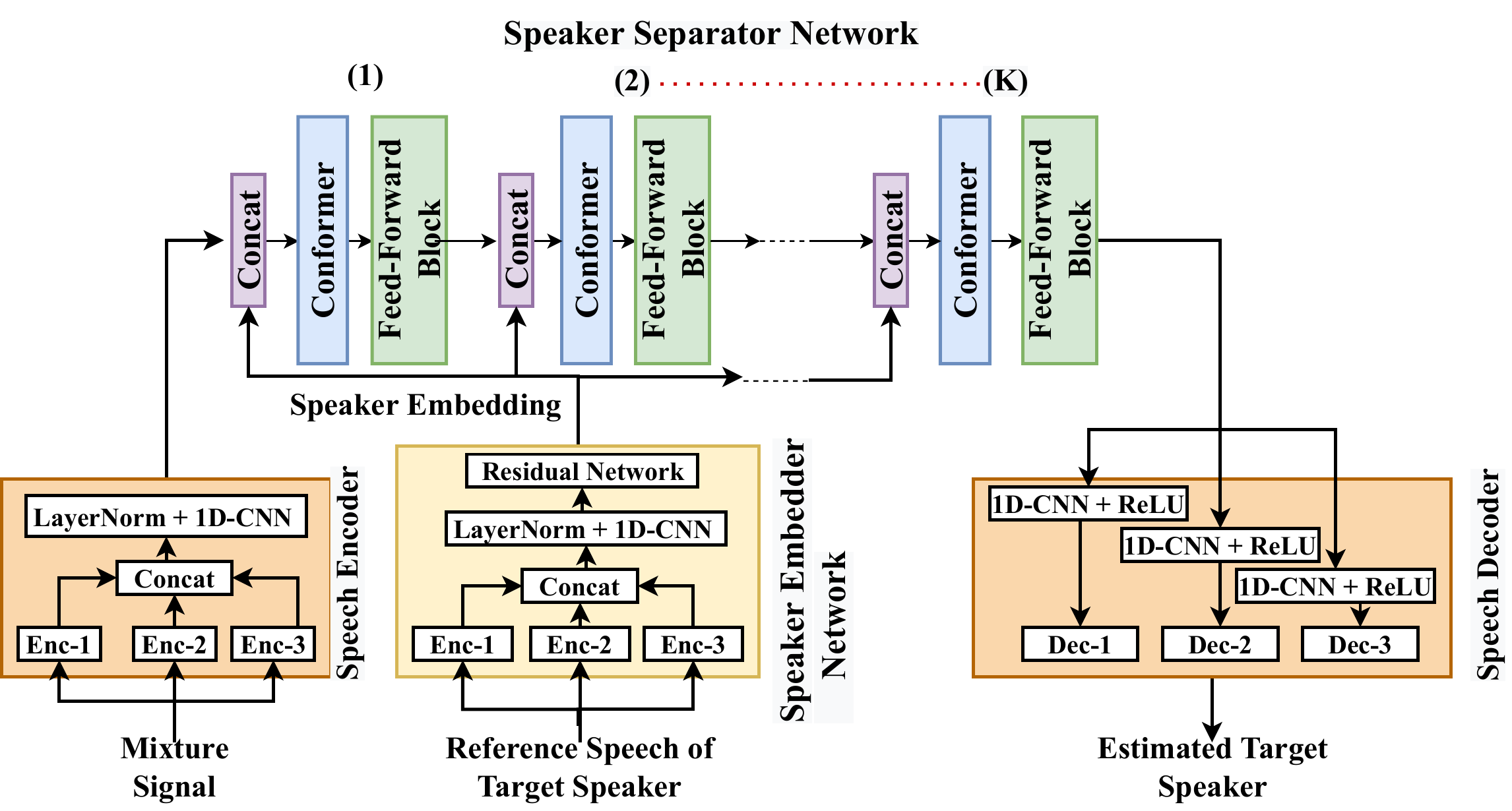}}
	\end{minipage}
	\caption{Proposed Conformer-FFN architecture, where {\it{K}} denotes the number of stacks of conformer and feed-forward blocks.}
	\label{fig:conf}
\end{figure}

\begin{figure}[tb]
	\begin{minipage}[b]{0.45\linewidth}
		\centering
		\centerline{\includegraphics[width=4.0cm]{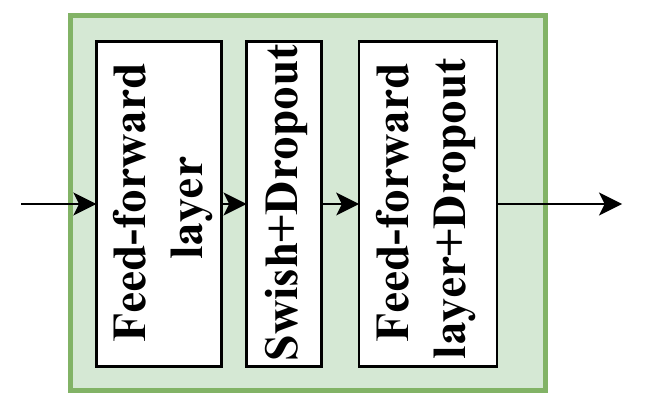}}
		\centerline{(a) Feed-Forward Block}\medskip
	\end{minipage}
	\begin{minipage}[b]{.45\linewidth}
		\centering
		\centerline{\includegraphics[width=3.5cm]{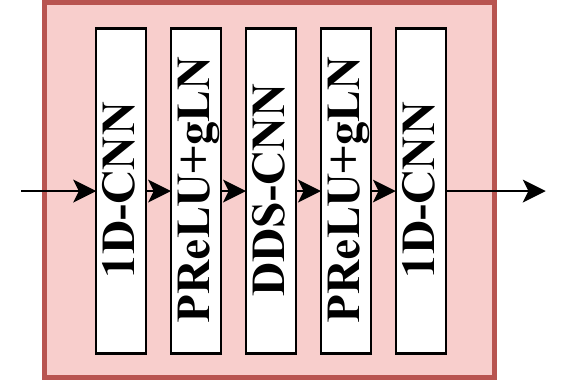}}
		\centerline{(b) TCN Block}\medskip
	\end{minipage}
	\caption{Feed-forward block and TCN block in speaker separator networks.}
	\label{fig:modules}
\end{figure}

Fig.~\ref{fig:conf} depicts the proposed Conformer-FFN architecture which is based on stacking conformer blocks and external feed-forward blocks. The motivation behind this architecture is to utilize both local and global context features using conformer blocks, while reducing the overall parameters using external feed-forward blocks. As proposed in \cite{gulati2020conformer} each conformer block consists of four different blocks: two feed-forward blocks, one multi-head self-attention block, and a convolutional block. The first feed-forward block is utilized before applying the multi-head attention. The convolutional block is utilized after the multi-head attention block just before the second feed-forward block. In the proposed architecture, each conformer block is followed by an external feed-forward block. The external feed-forward block consists of two feed-forward layers, a swish activation \cite{ramachandran2017searching}, and dropout layers (see Fig.~\ref{fig:modules}(a)). The output dimension of the external feed-forward block is half of the input dimension, leading to two advantages. First, we obtain a fixed dimensional input for all conformer blocks. Second, the overall number of parameters for the speaker separator network is reduced. The input to the first conformer block is the concatenation of the encoded features obtained from the mixture signal using the multi-scale encoder and the speaker embedding, while the inputs to the rest of the conformer blocks are the concatenation of the external feed-forward block output and the speaker embedding. Similarly to the TCN-based baseline \cite{ge2020spex+}, the speaker embedding is repeatedly concatenated along the feature dimension.

\begin{figure}[tb]
	\begin{minipage}[b]{1.0\linewidth}
		\centering
		\centerline{\includegraphics[width=8.4cm]{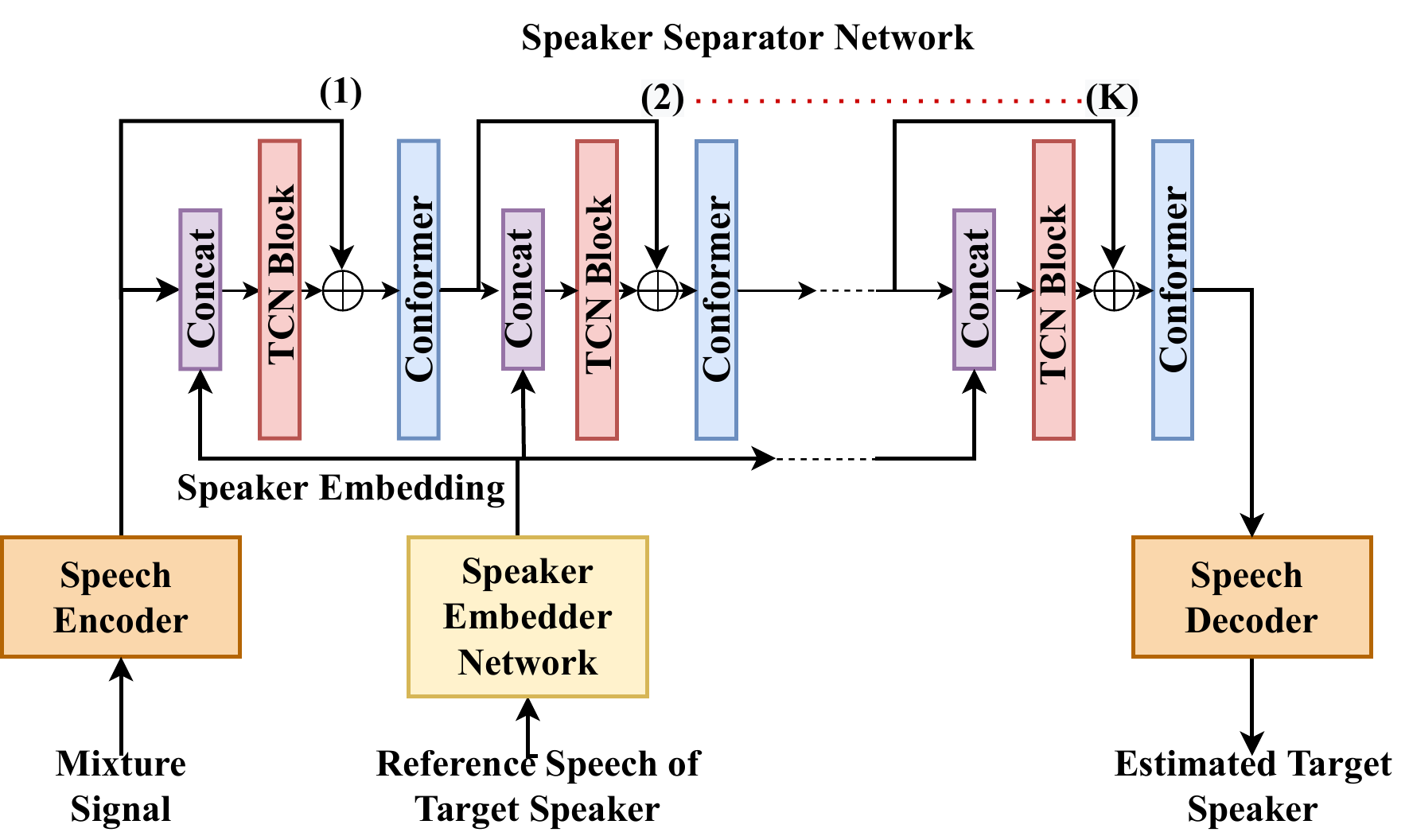}}
	\end{minipage}
	\caption{Proposed TCN-Conformer architecture, where {\it{K}} denotes the number of stacks of TCN and conformer blocks.}
	\label{fig:tcn-conf}
\end{figure}
\subsection{TCN-Conformer architecture}\label{sec:tcn-conf}
Fig.~\ref{fig:tcn-conf} depicts the proposed TCN-Conformer architecture, which is based on stacking TCN blocks and conformer blocks, i.e., each TCN block is followed by a conformer block. The motivation behind this architecture is to first utilize the best local context features using TCN blocks and then exploit both local and global context features using conformer blocks. Each TCN block \cite{luo2019conv} (see Fig.~\ref{fig:modules}(b)) consists of two 1-dimensional convolutional (1D-CNN) layers, two PReLU activations with global layer normalization (gln), and one dilated depth-wise separable convolutional layer (DDS-CNN). The conformer block consists of the same architecture as discussed in Section \ref{sec:conf}. The input to the first TCN block is the concatenation of the encoded features obtained from the mixture signal using the multi-scale encoder and the speaker embedding estimated using the speaker embedder network, while the inputs to the rest of TCN blocks are the concatenation of the conformer block output and speaker embedding. Similarly to the TCN-based baseline \cite{ge2020spex+}, the speaker embedding is repeatedly concatenated along the feature dimension.

\section{Experiments} \label{sec:exp}
In this section, we discuss the dataset for training and testing, the used parameters and hyper-parameters, and the training procedure of the considered target speaker extraction systems.

\subsection{Dataset}\label{sec:data}
We have simulated different datasets for 2-speaker mixtures (2-mix), 3-speaker mixtures (3-mix), and noisy mixtures of 2-speakers (noisy-mix) from the WSJ0 corpus \cite{hershey2016deep} and WHAM \cite{wichern2019wham} at a sampling rate of $16$ kHz. The subset $si\_tr\_s$ is utilized to create the training and development set, while the subsets $si\_dt\_05$ and $si\_et\_05$ from the WSJ0 corpus are utilized to create the test set. The test set contains completely different speakers than the training and developement sets. To create the 2-mix dataset, two different speakers are randomly chosen and mixed at an SNR between $0$ and $5$ dB, where the first speaker is regarded as the target speaker and the second speaker as the interfering speaker. A different utterance of the target speaker is selected as the reference speech to obtain the speaker embedding. We follow the same procedure to create the 3-mix dataset, where both interfering speakers have the same power, and the mixture with the target speaker is simulated at an SNR between $0$ and $5$ dB. Similarly, we create the noisy-mix dataset, where the target and interfering speakers are selected from the WSJ0 corpus and the noise is selected from the training, development and test set of WHAM. The noisy-mix data are simulated utilizing the official scripts of WHAM data simulation for 2-speaker mixtures. All together we have $47926$ utterances for training, $12792$ utterances for development, and $7478$ utterances for testing for 2-mix, 3-mix, and noisy-mix. 


\subsection{Training settings}\label{sec:setup}
We have used the same speaker embedder network as the baseline system \cite{ge2020spex+} to generate fixed $256$-dimensional embedding from the reference speech of the target speaker. The embedder network utilizes a ResNet architecture consisting of $3$ residual blocks. The input and output dimensions of the residual blocks are fixed to ($256$, $256$), ($256$, $512$), and ($512$, $512$) respectively. Each residual block consists of two 1D-CNN layers with a kernel size of $1$, where each CNN layer is followed with a batch-normalization and a PReLU activation function. A skip connection is used between the input and the second batch-normalization output, while a 1D max-pooling with a kernel size of $3$ is used as the output layer of each residual block.
As the baseline system, we have retrained the TCN-based target speaker extraction system proposed in \cite{ge2020spex+}. Each TCN block used in this work has similar hyper-parameter settings as the first TCN block utilized in \cite{ge2020spex+}. The input and convolutional size of the TCN block are fixed to $512$, and the kernel size is fixed to $3$. Each conformer block utilizes $8$-head attention, while the convolutional kernel size is fixed to $31$. The output of the first convolution layer is expanded with a factor of $3$ in each block, while the output of the linear layer is set to be $4$ times as input size. The external feed-forward blocks in the proposed Conformer-FFN architecture (see Fig.~\ref{fig:conf}) utilize an input dimension of $512$ and an output dimension of $256$ as well as a swish activation function (like the feed-forward block utilized in the conformer block).

To train all considered systems, we have utilized a multi-task objective function, namely multi-scale scale-invariant signal-to-noise ratio (SI-SNR) for the speaker separator network and cross-entropy for the speaker embedder network. Each system has been trained for $4$-seconds long segments for $150$ epochs with an early stopping criterion of $6$ epochs using the ADAM optimizer \cite{kingma2014adam}. The three filter lengths for the multi-scale encoder and decoder are set as $2.5$ ms, $10$ ms, and $20$ ms. The proposed Conformer-FFN and TCN-Conformer have been trained for {\it{$K\in\{1,3,4\}$}} number of stacks.

\section{Results and Discussion} \label{sec:results}
We first train the TCN-based baseline system and proposed Conformer-FFN and TCN-Conformer systems ({\it{$K=3$}} stacks) using only the 2-mix dataset. We then train all considered systems using all datasets (2-mix, 3-mix, noisy-mix) together. For both training conditions, we evaluate the performance for the 2-mix, 3-mix and noisy-mix test set individually. As performance measure, we use SI-SDR (dB) \cite{le2019sdr}, which is considered to be more robust than SDR \cite{vincent2006performance} for single-channel speaker extraction.

For all considered target speaker extraction systems trained only with the 2-mix dataset, Table \ref{tab:results-baseline} shows the mean SI-SDR values obtained on the 2-mix, 3-mix, and noisy-mix test sets. First, it can be observed that the proposed TCN-Conformer system improves the speaker extraction performance by $0.70$ dB for 2-mix, $0.38$ dB for 3-mix, and $2.06$ dB for noisy-mix compared to the TCN-based baseline system. Second, it can be observed that the proposed Conformer-FFN system achieves a lower performance than the baseline system for all test sets. Third, it can be observed that all considered systems trained only with the 2-mix dataset achieve high performance for the 2-mix test set, but perform poorly when either an additional interfering speaker (3-mix) or additional background noise (noisy-mix) is added.

\begin{table}[t]
	\centering
	\begin{tabular*}{0.91\columnwidth}{lcccc}
	   \hline
	    \textbf{Systems} &\textbf{{\it{K}}} & \textbf{2-mix} & \textbf{3-mix} & \textbf{noisy-mix}\\
		\noalign{\hrule height 1pt}
		Input mixture &- & 2.51 & -1.27 & -3.21 \\
		\hline
		baseline \cite{ge2020spex+} &- & 16.15 & 4.18 & -2.30\\ 
		\hline
		Conformer-FFN &3 & 15.60 & 4.08 & -3.64 \\
		\hline
            TCN-Conformer &3 & \textbf{16.85} & \textbf{4.56} & \textbf{-0.24}\\
		\noalign{\hrule height 1pt}
	\end{tabular*}
	\caption{SI-SDR (dB) for the input mixture, baseline system and proposed Conformer-FFN and TCN-Conformer systems trained only with the 2-mix dataset.}
	\label{tab:results-baseline}%
\end{table}

For all considered target speaker extraction systems trained with all datasets together, Table \ref{tab:results-proposed} shows the mean SI-SDR values obtained on the 2-mix, 3-mix, and noisy-mix test sets. First, it can be observed that for all mixtures all systems achieve a significant performance improvement with respect to the input mixture. Second, it can be observed that the performance for both proposed systems improves as the number of stacks {\it{$K$}} increases. Third, it can be observed that the best performing Conformer-FFN system ({\it{$K=4$}}) does not outperform the baseline system but the best performing TCN-Conformer system ({\it{$K=4$}}) improves the speaker extraction performance by $2.64$ dB for 2-mix, $2.27$ dB for 3-mix, and $1.40$ dB for noisy-mix compared to the baseline system.

\begin{table}[t]
	\centering
	\begin{tabular*}{0.91\columnwidth}{lcccc}
	   \hline
	    \textbf{Systems} & \textbf{\it{K}} & \textbf{2-mix} & \textbf{3-mix} & \textbf{noisy-mix}\\
		\noalign{\hrule height 1pt}
		Input mixture &- & 2.51 & -1.27 & -3.21 \\
		\hline
		baseline \cite{ge2020spex+} & - & 14.87 & 8.43 & 7.92\\
		\hline
        Conformer-FFN & 1 & 11.99 & 6.34 & 6.30\\
        \hline
        Conformer-FFN & 3 & 13.03 & 7.09 & 7.08\\
        \hline
        Conformer-FFN & 4 & 14.07 & 7.67 & 7.56\\
		\hline
		TCN-Conformer & 1 & 12.34 & 7.12 & 6.85\\ 
        \hline
        TCN-Conformer & 3 & \textbf{15.47} & \textbf{9.21} & \textbf{8.87}\\ 
        \hline
        TCN-Conformer & 4 & \textbf{17.51} & \textbf{10.70} & \textbf{9.32}\\
		\noalign{\hrule height 1pt}
	\end{tabular*}
	\caption{SI-SDR (dB) for the input mixture, baseline system and proposed Conformer-FFN and TCN-Conformer systems trained with all considered datasets (2-mix, 3-mix, noisy-mix) together.}
	\label{tab:results-proposed}%
\end{table}

\section{Conclusion} \label{sec:last}
In this paper, we explored the benefits of using conformer-based architectures for the separator network in time-domain speaker-conditioning target speaker extraction systems. Aiming at exploiting both local as well as global context features, we proposed the Conformer-FFN architecture, stacking conformer blocks and external feed-forward blocks and the TCN-Conformer architecture, stacking TCN blocks and conformer blocks. Experimental results for different mixtures show that the proposed TCN-Conformer outperforms the proposed Conformer-FFN system and a TCN-based baseline system. In future work, we will investigate the effect of noisy auxiliary information on the performance of target speaker extraction systems.

\section{References} \label{sec:ref}
\small
\bibliographystyle{IEEEtran}
\bibliography{refs}

\end{document}